\documentclass[a4paper,fleqn,usenatbib]{mnras}

\usepackage{mathptmx}
\usepackage{graphics,graphicx}	
\usepackage{amsmath}	
\usepackage{amssymb}	
\usepackage{float}
\usepackage[T1]{fontenc}
\usepackage{ae,aecompl}
\usepackage{amsmath} 
\newcommand{\angstrom}{\text{\normalfont\AA}}

\title[GRS 1734-292]{Broadband X-ray spectral analysis of the Seyfert 1 galaxy  GRS 1734-292}

\author[A. Tortosa et al.]{A. Tortosa$^{1}$\thanks{E-mail: tortosa@fis.uniroma3.it}, A. Marinucci$^{1}$, G. Matt$^{1}$, S. Bianchi$^{1}$, F. La Franca$^{1}$, D. R. Ballantyne$^{2}$,
\newauthor  P. G. Boorman$^{3}$, A.C. Fabian$^{4}$, D. Farrah$^{5}$, F. Fuerst$^{6}$, P. Gandhi$^{3}$, F. A. Harrison$^{6}$,
\newauthor  M. J. Koss$^{7}$, C. Ricci$^{8,9}$,  D. Stern$^{10}$,  F. Ursini$^{1,11,12}$, D. J. Walton$^{10,6}$. \\
$^{1}$Dipartimento di Matematica e Fisica, Universit\`{a} degli Studi Roma Tre, via della Vasca Navale 84, 00146 Roma, Italy. \\
$^{2}$ Center for Relativistic Astrophysics, School of Physics, Georgia Institute of Technology, 837 State Street, Atlanta, GA 30332- 0430, USA.\\
$^{3}$ School of Physics and Astronomy, University of Southampton, Highfield, Southampton SO17 1BJ, UK.\\
$^{4}$ Institute of Astronomy, Madingley Road, Cambridge CB3 OHA.\\
$^{5}$ Department of Physics, Virginia Tech, Blacksburg, VA 24061, USA.\\
$^{6}$ Space Radiation Laboratory, California Institute of Technology, Pasadena, CA 91125, USA.\\
$^{7}$ Institute for Astronomy, Department of Physics, ETH Zurich, Wolfgang-Pauli-Strasse 27, CH-8093 Zurich, Switzerland.\\
$^{8}$ Instituto de Astrof\'{i}sica, Facultad de F\'{i}sica, Pontificia Universidad Cat\'{o}lica de Chile, Casilla 306, Santiago 22, Chile.\\
$^{9}$ EMBIGGEN Anillo, Concepci\'{o}n, Chile.\\
$^{10}$ Jet Propulsion Laboratory, California Institute of Technology, Pasadena, CA 91109, USA.\\
$^{11}$ Univ. Grenoble Alpes, CNRS, IPAG, F-38000 Grenoble, France\\
$^{12}$ CNRS, IPAG, F-38000 Grenoble, France.
}

\date{Accepted XXX. Received YYY; in original form ZZZ}

\pubyear{2016}

\begin{document}
\label{firstpage}
\pagerange{\pageref{firstpage}--\pageref{lastpage}}
\maketitle

\begin{abstract}
We discuss the broadband X-ray spectrum of GRS 1734-292 obtained from non-simultaneous \textit{XMM-Newton} and \textit{NuSTAR} observations, performed in 2009 and 2014, respectively.\\
GRS1734-292 is a Seyfert 1 galaxy, located near the Galactic plane at $z=0.0214$.
The \textit{NuSTAR} spectrum ($3-80$ keV) is dominated by a primary power-law continuum with $\Gamma=1.65 \pm 0.05$ and a high-energy cutoff $E_c=53^{+11}_{-8}$ keV, one of the lowest measured by \textit{NuSTAR} in a Seyfert galaxy. Comptonization models show a temperature of the coronal plasma of $kT_e=11.9^{+1.2}_{-0.9}$ keV and an optical depth, assuming a slab geometry, $\tau=2.98^{+0.16}_{-0.19}$ or a similar temperature and  $\tau=6.7^{+0.3}_{-0.4}$ assuming a spherical geometry. The 2009 \textit{XMM-Newton} spectrum is well described by a flatter intrinsic continuum ($\Gamma=1.47^{+0.07}_{-0.03}$) and one absorption line due to Fe\textsc{XXV} K$\alpha$ produced by a warm absorber. Both data sets show a modest iron K$\alpha$ emission line at $6.4$ keV and the associated Compton reflection, due to reprocessing from neutral circumnuclear material. 
\end{abstract}

\begin{keywords}
galaxies: active -- galaxies: Seyfert -- X-rays: galaxies -- galaxies: individual: GRS 1734-292
\end{keywords}



\section{INTRODUCTION}
\label{intro}
The primary X-ray emission in Active Galactic Nuclei (AGN) is believed to be produced in a compact hot region, located close to the supermassive black hole, and composed of a plasma of hot electrons: the corona (\cite{HeM1993}). The optical/UV disk photons are scattered up to the X-ray band due to the inverse Compton effect in this hot corona. The Comptonization leads to a power-law continuum, extending up to energies determined by the plasma electron temperature ($kT_e$) (\cite{RibLig}), where a cutoff occurs. The power-law index is directly related to the temperature and optical depth of the plasma of hot electrons responsible for the power-law emission while the cutoff energy is mainly related to the electron temperature.\\ 
Pre-{\it NuSTAR} (\textit{Nuclear Spectroscopic Telescope Array}: \cite{Harrison}) measurements of cutoff energies ranged between 50 and 300 keV (e.g. \cite{Perola02}, \cite{Dadina}, \cite{Malizia2014}). {\it NuSTAR}'s high sensitivity in hard X-rays, allowing for the first time source-dominated observations of Seyfert galaxies above 10 keV, has recently led to high-energy cutoff measurements from 100 keV to more than 350 keV for several nearby Seyfert galaxies (see: \cite{ic}, \cite{swiftj}, \cite{ic}, \cite{3c}, \cite{52315}, \cite{5506}), plus a number of significant lower limits (\cite{Fabian}).\\
The nearby ($z=0.0214$, corresponding to a distance of $87$ Mpc) Seyfert Galaxy GRS 1734-292 is a good candidate for such measurements. With an X-ray luminosity approaching $\sim 10^{44}$ erg s$^{-1}$ in the 0.5-4.5 keV energy band (\cite{Marti1998}), it is one of the most luminous AGNs within 100 Mpc (\cite{Piccinotti1982}; \cite{SazoRev2004}). \\
GRS 1734-292 was originally discovered by the ART-P telescope aboard the \textit{GRANAT} satellite (\cite{Pavlinsky1992}) and is located only $1.8^{\circ}$ from the Galactic Centre. The spectrum between $4-20$ keV was well described by a power-law with a photon index $\Gamma \sim 2$  and a total hydrogen column density in excess of $10^{22}$ cm$^{-2}$. These characteristics, with the inferred X-ray  luminosity $\sim 10^{36}$erg s$^{-1}$ assuming the Galactic Center distance, were consistent with the source being a Galactic X-ray binary. \cite{Marti98} revealed that the optical spectrum of GRS 1734-292 is dominated by strong and very broad emission from blended H and [NII] lines, but also other emission lines, such as [OI], [OII] and [SII], all at a redshift of 0.0214. Moreover, the radio, infrared and optical counterparts of GRS 1734-292 are all consistent with a Seyfert 1 galaxy. In particular,  the radio counterpart is a double-sided synchrotron jet of $5$ arcsec extent. At the distance of $87$ Mpc, this corresponds to a size of 2 kpc. With a radio luminosity of $L_{\rm rad} \simeq 7 \times 10^{39}$erg s$^{-1}$ in the $0.1-100$ GHz band and an X-ray luminosity $L_{X} \simeq 1 \times 10^{44}$ erg s$^{-1}$ in the $0.5-4.5$ keV band, GRS 1734-292 is a radio-quiet AGN (\cite{Laor2008}).\\
The hard X-ray spectrum of GRS 1734-292 was measured for the first time with the IBIS telescope onboard the \textit{INTEGRAL} observatory (\cite{Sazonov2004}). Afterwards it was also analyzed by \cite{molina13}. The composite X-ray ($2-200$ keV) spectrum with the\textit{ ASCA}/GIS observation at $2-10$ keV (\cite{Sakano2002}) is typical of Seyfert galaxies, well described by a power-law of $\Gamma \sim 1.8 $ modified by Compton reflection at $10-100$ keV and an exponential cutoff at  $E_{\rm c}>100-200$ keV.\\
GRS 1734-292 was detected also in 70 months of observations by the BAT hard X-ray detector (\cite{Barthelmy}) on the \textit{Swift} gamma-ray burst observatory (\cite{Gehrels}). The spectral analysis (\cite{Baumgartner2011}) showed a power law with a photon index $\Gamma \sim 2.18 \pm 0.07 $ and a luminosity of L$ \sim 1-2 \times 10^{44} $ erg s$^{-1}$ in the $14-195$ keV band.\\
\cite{guainazzi11} analyzed the \textit{XMM-Newton} observation in their GREDOS (General Relativity Effects Detected in Obscured Sources) sample and found that the spectrum is well described by a power-law with a rather flat spectral index $\Gamma =1.41^{+0.01}_{-0.02}$. They found a hydrogen column density of $N_{\rm H}=1.41 \pm 0.02 \times 10^{22}$cm$^{-2}$. From the spectral analysis of the simultaneous \textit{XMM-Newton} and \textit{INTEGRAL}/IBIS-\textit{Swift}/BAT observations, \cite{Malizia14} found a primary continuum with a power-law index $\Gamma = 1.55^{+0.15}_{-0.08}$ and a cutoff energy $E_c=58^{+24}_{-7}$ keV.\\
This work focuses on investigating the broad band X-ray spectrum of GRS 1734-292 and in particular the physical properties of the corona. In Sect.\ref{obs} we discuss the \textit{XMM-Newton} and \textit{NuSTAR} observations and data reduction. In Sect.\ref{analysis} we present a re-analysis of the 2009 \textit{XMM-Newton} observation together with a new \textit{NuSTAR} observation of GRS 1734-292. We discuss our results and summarize our conclusions in Sect.\ref{concl}. In Appendix \ref{optical}, a measure of the width of the broad H$\alpha$ $\lambda6563$ component, to infer the black hole mass via an updated virial-based, single-epoch relation, is presented.

\section{OBSERVATIONS \& DATA REDUCTION}
\label{obs}
\subsection{\textit{NuSTAR}}
GRS 1734-292 was observed by {\it NuSTAR}  with its two coaligned X-ray telescopes Focal Plane Modules A and B (FPMA and FPMB, respectively)  on 2014 September 16, for a total elapsed time of 43 ks. The Level 1 data products were processed with the {\it NuSTAR} Data Analysis Software (NuSTARDAS) package (v. 1.3.0). Cleaned event files (level 2 data products) were produced and calibrated using standard filtering criteria with the {\sc nupipeline} task and the latest calibration files available in the {\it NuSTAR} calibration database (CALDB 20150316). The extraction radii of the circular region for source and background spectra were 1.5 arcmin each; there is no other bright X-ray source within 1.5 arcmin from GRS 1734 and no other sources were present in the background region. The net exposure times after this process were 20.3 ks for both FPMA and B. The two spectra were binned in order to over-sample the instrumental resolution by at least a factor of 2.5 and to have a Signal-to-Noise Ratio (SNR) greater than 3 in each spectral channel.
\begin{figure}
\includegraphics[width=0.7\columnwidth, angle=-90]{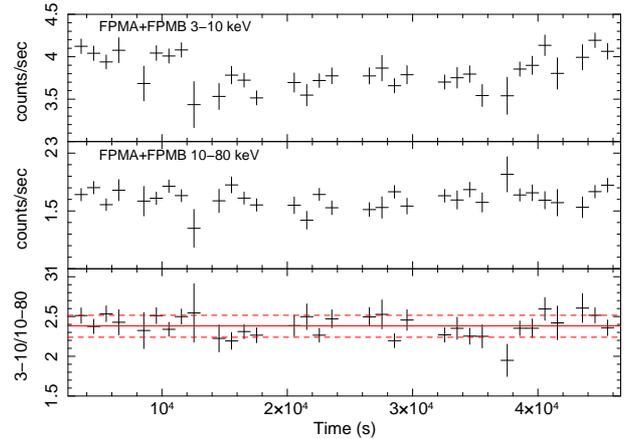}
\caption{Top panel: \textit{NuSTAR} FPMA+B light curve in the 3-10 keV energy band; middle panel: \textit{NuSTAR} FPMA+B light curve in the 10-80 keV energy band; bottom panel: ratio between 3-10 keV and 10-80 keV \textit{NuSTAR} light curves, the red solid and dashed lines indicate the mean and standard deviation, respectively.}
 \label{lc}
\end{figure}
Since no spectral variation (less than 10\%) is found in the ratio between the $3-10$ and $10-80$ keV count rates (see Figure \ref{lc}) we decided to use time-averaged spectra.\\
 Since GRS 1734-292 lies very low on the galactic plane, the {\it NuSTAR} observation is moderately affected by stray light, due to sources off the field of view. This effect is more significant in the FPMB detector: a $50\pm2\%$ increase in the background count rate is observed below $7$ keV, with respect to the FPMA. We tried to extract the background spectra from different regions and no differences are found. Since the point source falls within the stray light region in both the detectors and the background is hence properly subtracted, this effect is not relevant to our data analysis. As a last check, we verified that no spectral difference arise between the two {\it NuSTAR} background subtracted spectra: they perfectly agree within cross-calibration uncertainties.
\subsection{\textit{XMM-Newton}}
GRS 1734-292 was also observed with \textit{XMM-Newton} on 2009 February 26 with the EPIC CCD cameras, which are comprised of the pn detector  (\cite{struder2001}) and the two MOS units (\cite{turner2001}).The camera were operated in large window and thin filter mode, for a total elapsed time of 18 ks. The extraction radii and the optimal time cuts for flaring particle background were computed with SAS 15 (\cite{Gabriel2004}) via an iterative process which maximizes the SNR, similar to the approach described in \cite{Piconcelli2004}. The resulting optimal extraction radius was 40 arcsec and the background spectra were extracted from source-free circular regions with radii of $\sim 50$ arcsec for both the EPIC and the two MOS. EPIC spectra had a net exposure time of $13$ ks, the MOS spectra had both a net exposure time of $15$ ks. EPIC and MOS spectra were binned in order to over-sample the instrumental resolution by at least a factor of three and to have no less than $30$ counts in each background-subtracted spectral channel.
Data from the MOS detectors are not included in our analysis unless stated otherwise.\\

\section{SPECTRAL ANALYSIS}
\label{analysis}
The spectral analysis has been performed with the \textsc{xspec 12.9.0} software package (\cite{xspec}). Throughout the paper, 
errors correspond to the 90\% confidence level for one interesting parameter ($\Delta\chi^2=2.7$), if not stated otherwise. 
The cosmological parameters $H_0=70$ km s$^{-1}$ Mpc$^{-1}$, $\Omega_\Lambda=0.73$ and $\Omega_{\rm\,m}=0.27$ are adopted.

\subsection{Re-analysis of the \textit{XMM-Newton} data}
\begin{figure}
\includegraphics[width=0.7\columnwidth, angle=-90]{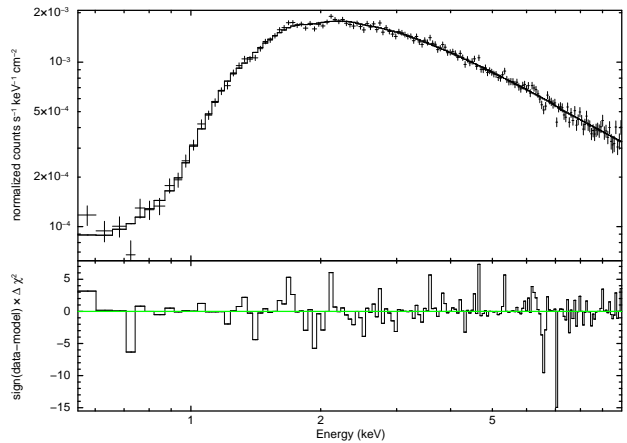}
\caption{Data and residuals for the \textit{XMM-Newton}  spectrum when no Gaussian lines are included in the model.}
 \label{chi_xmm}
\end{figure}

We started our data analysis by fitting the $0.5-10$ keV \textit{XMM-Newton} spectrum with a model \footnote{\textsc{xspec} model: TBabs * zwabs * powerlaw} composed of a power law absorbed by the Galactic column density $N_{\rm H}=7.57\times10^{21}$ cm$^{-2}$, as derived from HI maps (\cite{heasarc}), and an additional intrinsic absorber at the redshift of the source, found to be $0.84 \pm 0.03 \times 10^{22}$ cm$^{-2} $ . 
This yelded a poor fit with $\chi^{2}=242$ for $163$ degrees of freedom (d.o.f.). Since the data reveal a slight excess at energies $<1$ keV, we added a \textsc{mekal} thermal plasma component\footnote{\textsc{xspec} model: TBabs * (mekal + zwabs * powerlaw)} (\cite{mekal}), absorbed only by Galactic gas. We fixed the chemical abundance of metals to solar. The $\chi^{2}$ is $209$ for $161$ d.o.f. for a thermal plasma temperature of $kT=0.14^{+0.22}_{-0.05}$ keV and a 0.5-2 keV luminosity of $\sim 2.7 \pm 0.7 \times$10$^{42}$ erg s$^{-1}$. Some residuals are however evident around $6-7.5$ keV (see Figure \ref{chi_xmm}). 
Therefore we added a narrow Gaussian line\footnote{\textsc{xspec} model: TBabs * (mekal + zwabs * (zgauss + powerlaw))} at $6.4$ keV, corresponding to the neutral iron K$\alpha$ emission line, which is a typical feature in Seyfert galaxies \citep{Nandra94}. We found the centroid value of the line to be $6.36 \pm 0.07$ keV and the fit slightly improved: $\chi^2=202$ for $159$ d.o.f., with a $\frac{\Delta \chi ^2}{\Delta \rm {d.o.f.} }=3.5$ and a null hypothesis probability of $4.7 \times 10^{-2}$ according to the \textit{F}-test. The iron K$\alpha$ emission line shows a flux of $1.4 \pm 0.8 \times 10^{-5}$ ph cm$^{-2}$ s$^{-1}$ and an equivalent width of $20 \pm 13$ eV. We then added an absorption Gaussian line\footnote{\textsc{xspec} model: TBabs*(mekal + zwabs*(zgauss + zgauss + powerlaw)), suggested by the presence of negative residuals around $6.7$ keV. We found a centroid value for this line of $6.69 \pm 0.05$ keV; the $\chi^2$ is $187$ for $157$ d.o.f., with a $\frac{\Delta \chi ^2}{\Delta \rm {d.o.f.} }=7.5$ and a null hypothesis probability of $1.5 \times 10^{-3}$ according to the \textit{F}-test }\footnote{In principle, the \textit{F}-test is not a reliable test for the significance of emission or absorption lines, but it can be used if their normalizations are allowed to be negative and positive (\cite{protassov}).}. The flux and the line equivalent width of this absorption line were $ 2.1 \pm 0.8 \times 10^{-5}$ ph cm$^{-2}$ s$^{-1}$ and $31 \pm 12$ eV, respectively. The centroid energy of this absorption line is consistent with the K-shell transition of Fe\textsc{XXV} ions. We tried to fit this component with a warm-absorber (WA) model\footnote{\textsc{xspec} model: TBabs * (mekal + mtable$\lbrace$cloudy.fits$\rbrace$ * zwabs * (zgauss + powerlaw))}, using an ad-hoc table produced with the photo-ionization code \textsc{cloudy c13.03} (most recently described by \cite{ferland}). We found a ionization parameter of $\xi_i=1778.3^{+2.7}_{-1.6}$ erg cm s$^{-1}$ and a column density $N_{\rm H}=5.01 \pm 3.2 \times 10^{22}$ cm$^{-2}$. The upper limit to the velocity of the emitting material is $v_{max}=5300$ km s$^{-1}$. The $\chi^2$ is $186$ for $157$ d.o.f.. To check the possibility that the absorption features are due to a collisional ionized gas we fitted the data using the hotabs model in the warmabs code \citep{kallman} instead of the warm-absorber model.The fit is equivalent from a statistical point of view. We found a temperature of the gas of $0.85 \pm 0.12$ keV and a column density of $7.1 \pm 5.0 \times 10^{21}$ cm$^{-2}$. Further residuals around $7.2$ keV suggested to add another Gaussian absorption line\footnote{\textsc{xspec} model: TBabs * (mekal + mtable$\lbrace$cloudy.fits$\rbrace$ * zwabs * (zgauss + zgauss + powerlaw)) }. The inclusion of this component leads to a $\chi^{2}/$d.o.f.$=178/155=1.14$ (\textit{F}-test null hypothesis probability $2.2 \times 10^{-2}$). The fit gives a centroid energy of $7.19^{+0.07}_{-0.09}$ keV, with a flux of $1.3 \pm 0.8 \times 10^{-5}$ ph cm$^{-2}$ s$^{-1}$ and a line equivalent width (EW) of $28 \pm 14$ eV. An absorption line with this centroid energy is possibly a blue-shifted line associated with the transition of Fe\textsc{XXVI} ions (rest frame energy: $6.966$ keV) produced by a material with a velocity of $9500$ km s$^{-1}$ $\simeq0.03$c. This is the lower limit of the range of velocities for Ultra Fast Outflows (\cite{Tombesi}). To verify the presence of this line we fitted the pn and the MOS spectra simultaneously with the same model. We tied all of the MOS parameters to the pn values. The normalizations of the two Gaussian lines and the normalization of the power-law of the MOS spectra are tied together but are free to vary. The $\chi^2$ of the fit is  $461$ for $427$ degrees of freedom. The fit with the MOS data confirms the presence of the absorption line due to Fe\textsc{XXV} K$\alpha$ ions produced by a warm absorber, but not that at 7.2 keV. In fact, the upper limit to the flux of the latter line is $3.16 \times 10^{-6}$ ph cm$^{-2}$ s$^{-1}$ at 90\% confidence level. In the following fits, therefore, this line will not be included.\\ 
We found that the photon index of the primary X-ray continuum is $\Gamma = 1.47 \pm 0.03$.
This\footnotemark[6] is our best fit and we will use it as the baseline model when adding the {\it NuSTAR} data.
\subsection{Adding \textit{NuSTAR} data.}
We started the analysis of the $3-80$ keV \textit{NuSTAR} (FPMA and FPMB) spectra fitting the data together with the \textit{XMM-Newton} best fit found previously. We left all the parameters, apart from the normalizations of the various components, tied to the \textit{XMM-Newton} best fitting parameters. The \textit{XMM-Newton} and the \textit{NuSTAR} FPMA calibration constants are fixed to $1.0$ (given the non-simulateneity of the two observations, any mismatch between the two instruments cannot be separated from intrinsic variations) while we left the \textit{NuSTAR} FPMB cross-calibration constant free to vary. The value found for the constant is $1.004$. The $\chi^2$ for this fit is $830$ for $544$ d.o.f..
\begin{figure}
\includegraphics[width=0.69\columnwidth, angle=-90]{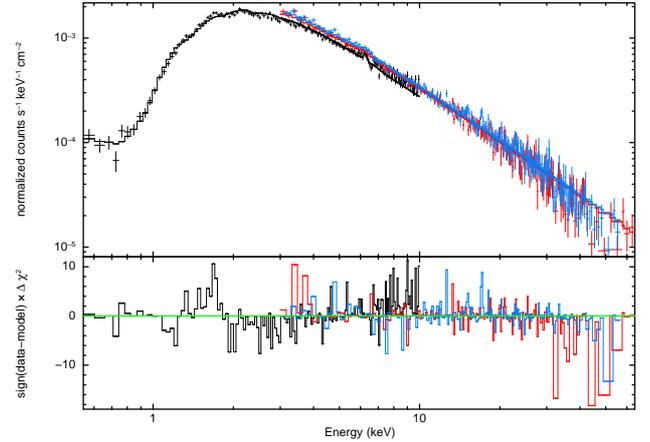}
\caption{Data, fit model (top panel) and residuals (bottom panel) for \textit{XMM-Newton} (black) and \textit{NuSTAR} FPMA (red) and FPMB (in blue) spectra when the parameters are all tied to the best fitting parameters from the \textit{XMM-Newton} spectral fit.}
\label{tiedgamma}
\end{figure}
The spectral slope shows a different trend for the power-law from the two observations (see Figure \ref{tiedgamma}) so we left the two photon indices, which are related to two different observations, free to vary. We kept tied the emission and absorption line centroid energies to the values found by \textit{XMM-Newton} due to the lower spectral resolution of \textit{NuSTAR}. We found that the \textit{NuSTAR} photon index is steeper than the \textit{XMM-Newton} one ($\Gamma = 1.65 \pm 0.05$). The fit leads to a $\chi^{2}/$d.o.f.$=662/543=1.22$. 
\begin{figure}
\includegraphics[width=0.69\columnwidth, angle=-90]{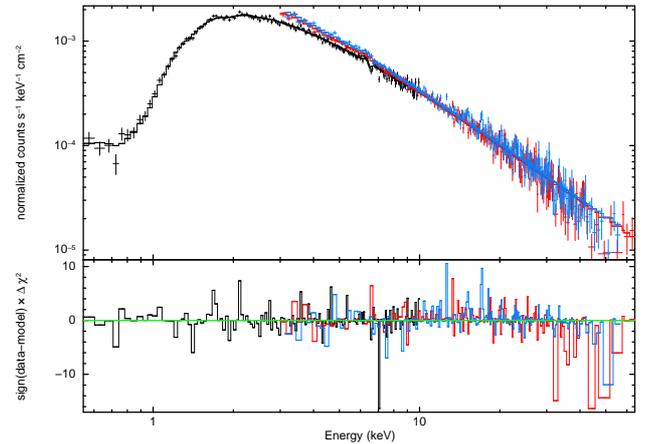}
\caption{Data, fit model (top panel) and residuals (bottom panel) for \textit{XMM-Newton} (black) and \textit{NuSTAR} FPMA (red) and FPMB (blue) spectra when the power law in the model is not corrected by a high energy cutoff. The photon indices of \textit{XMM} and \textit{NuSTAR} are left free to vary.}
 \label{nocutoff}
\end{figure}
Re-analyzing \textit{Swift}/BAT observation from the Swift BAT 70-Month Hard X-ray Survey \citep{nasa}, we found a photon index $\Gamma = 2.18 \pm 0.07$ consistent with (\cite{Baumgartner2011}). Adding a high energy cutoff, however, we found a flatter photon index $\Gamma = 1.8 \pm 0.3$ and an high energy cutoff value of $E_c = 110^{+300}_{-50}$ keV. The average \textit{Swift}/BAT flux is higher than the \textit{NuSTAR} one, which in turn is higher than \textit{XMM-Newton}'s one. The source therefore show the softer-when-brighter behaviour (\cite{shemmer}, \cite{sobolewska} ) which is typical for Seyfert Galaxies.\\
Back to the \textit{NuSTAR} data analysis, looking at the residuals above $\sim 40$ keV (see Figure \ref{nocutoff}) the presence of a high-energy cutoff is suggested, so we replaced the power law component with a power law corrected by a high energy exponential rolloff (\textsc{cutoffpl} model in \textsc{xspec})\footnote{\textsc{xspec} model: constant * TBabs * (mekal  + mtable$\lbrace$cloudy.fits$\rbrace$ * zwabs * (zgauss + cutoffpl))}. The fit improved significantly ($\chi^{2}/$d.o.f.$=556/541=1.1$); we found for the \textit{NuSTAR} spectra $\Gamma = 1.58 \pm 0.04$ with the cutoff energy $E_{\rm c}=60^{+17}_{-9}$ keV and $\Gamma = 1.40^{+0.06}_{-0.09}$ for the \textit{XMM-Newton} spectrum with a lower limit for the cutoff energy at $90$ keV.\\
We then included a cold reflection component in both the data sets, using the \textsc{pexrav} model (\cite{Magdziarz1995}) in \textsc{xspec}, to test for the presence of a Compton reflection continuum. We fixed all element abundances to solar values and fixed the inclination angle to the default value ($\cos \mathit{i} = 0.45$, $\mathit{i} \sim 60^{\circ}$). Because in the previous fit we found only a lower limit to the high-energy cutoff in the \textit{XMM-Newton} spectrum, for the sake of simplicity we fixed it to $1$ MeV. The model used in the fit is shown in Figure \ref{model}. Data and residuals are shown in Figure \ref{bestfit}, while the best fitting parameters are shown in Table \ref{fit_pexrav}. The photon index and high energy cutoff are now $\Gamma = 1.65 \pm 0.05$ and $E_{\rm c}=53^{+11}_{-8}$ keV. The reflection fraction $R$ is $0.48 \pm 0.22$. In the left panel of Figure \ref{fig_cont} the contour plot of the cutoff energy versus the photon index of the power law for the \textit{NuSTAR} observation is shown, while in the right panel we show the contour plot of the high energy cutoff versus the reflection fraction.\\ 
\begin{figure}
\includegraphics[width=0.8\columnwidth, angle=-90]{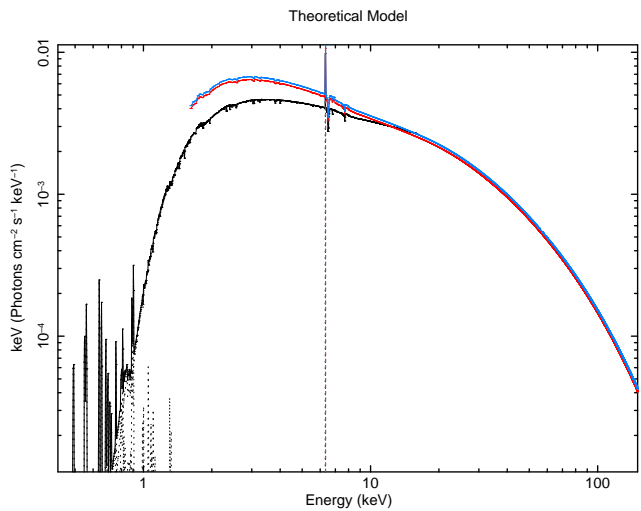}
\caption{Best fitting phenomenological model including the soft excess component, two narrow Gaussian lines, the WA and a cutoff power law reflected from neutral material (PEXRAV model), all absorbed by the Galactic column density and an intrinsic absorber. }
 \label{model}
\end{figure}
\begin{figure}
\includegraphics[width=0.69\columnwidth, angle=-90]{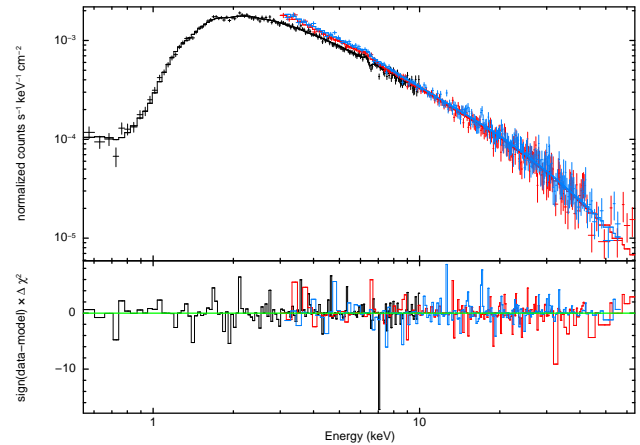}
\caption{Data and best fit model extrapolated from \textit{XMM-Newton} (black) and \textit{NuSTAR} FPMA (red) and FPMB (blue) spectra when model in Figure \ref{model} is used; see text for more details. Residuals are shown in lower panel}
 \label{bestfit}
\end{figure}
\begin{table}
\caption{Best fitting parameters for the phenomenological model including PEXRAV and two Gaussian lines (line 1 at $6.4$ keV and line 2 at $6.68$ keV, obtained from the \textit{XMM-Newton} spectrum). Errors are at $90 \%$ confidence levels. The $\chi^{2}$ / d.o.f. value is $580/ 540 = 1.07$.}
	\centering
	\begin{tabular}{ccc}
		\hline
		Parameter & \textit{XMM-Newton} & \textit{NuSTAR} \\
		\hline
		$N_{\rm H} (10^{22}$cm$^{-2})$ & $0.88 \pm 0.05$& $0.88^{**}$\\
		$\Gamma$ & $1.47^{+0.07}_{-0.03}$ & $1.65 \pm 0.05$ \\
		$E_{\rm c}$ (keV) & $1000^{*}$ & $53^{+11}_{-8}$ \\
		$R$ & $< 0.6$ & $0.48 \pm 0.22$ \\
		$F_{\rm 2-10}$ ($10^{-11}$ erg cm$^{-2}$ s$^{-1}$) & $5.12^{+0.15}_{-0.08} $ & $6.62^{+0.02}_{-0.08} $ \\
		$L_{\rm 2-10}$ ($10^{43}$ erg s$^{-1}$) & $5.23 \pm 0.03$ & $6.67 \pm 0.04$ \\
		$F_{\rm 10-80}$ ($10^{-10}$ erg cm$^{-2}$ s$^{-1}$) & - & $1.25 \pm 0.01 $ \\
		$L_{\rm 10-80}$ ($10^{44}$ erg s$^{-1}$) & - & $1.29 \pm 0.05$ \\
		$F_{\rm 1}$($10^{-5}$ ph/cm$^{2}$/s)&$1.37 \pm 0.84$&$3.93 \pm 1.91$\\
		EW$_{\rm 1}$ (eV) &$20 \pm 13 $& $50 \pm 31$ \\  
		$F_{\rm 2}$($10^{-5}$ ph/cm$^{2}$/s)&$2.06 \pm 0.77$&$0.75 \pm 1.86$\\
		EW$_{\rm 2}$ (eV) &$-31\pm 12 $& $< 28 $ \\ 
		\hline
	\end{tabular}
	\begin{flushleft}
	$^{*}$ fixed parameter.\\
	$^{**}$ tied parameter.
	\end{flushleft}
	\label{fit_pexrav}
\end{table}
\begin{figure*}
\includegraphics[width=0.7\columnwidth, angle=-90]{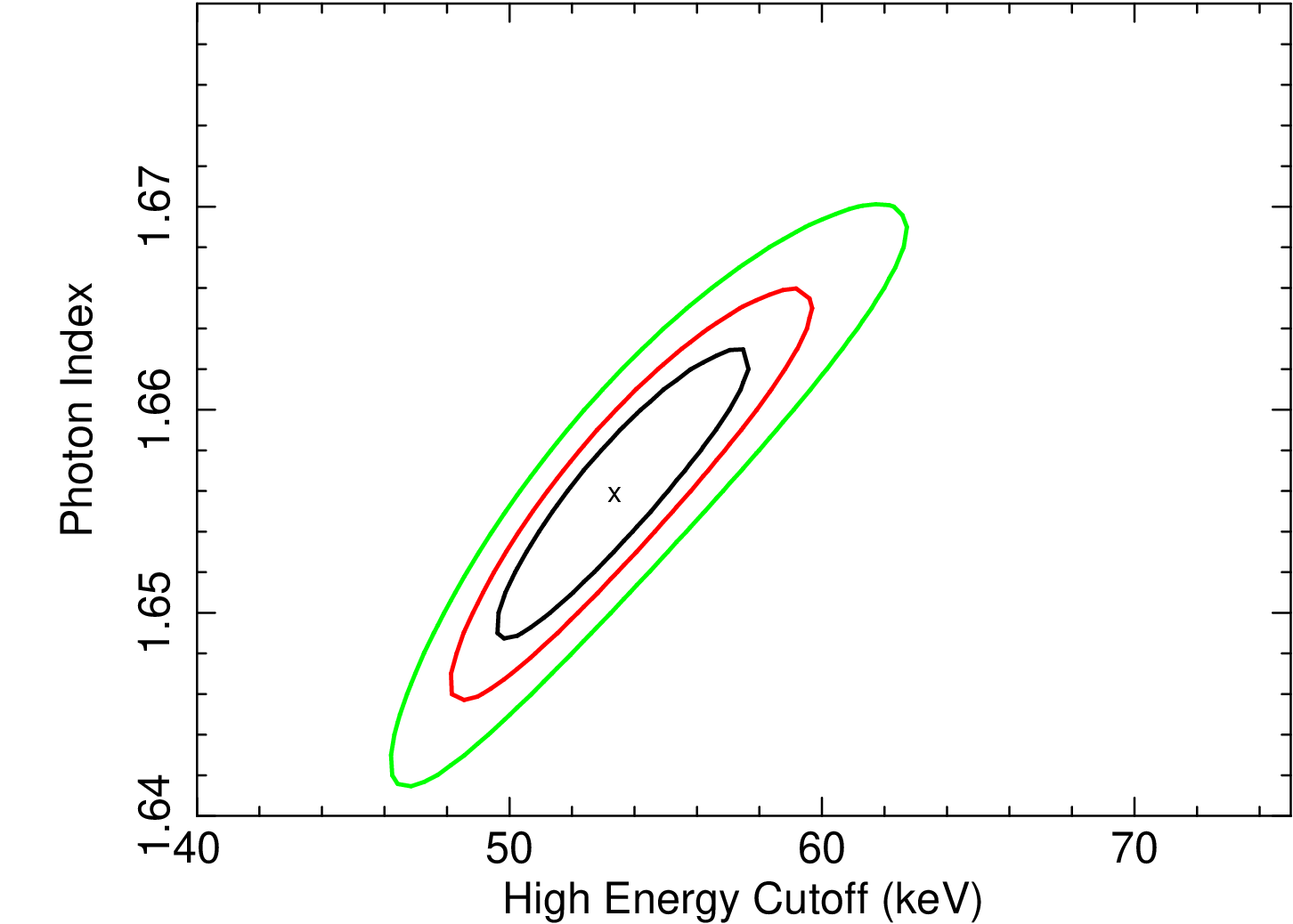}
\includegraphics[width=0.7\columnwidth, angle=-90]{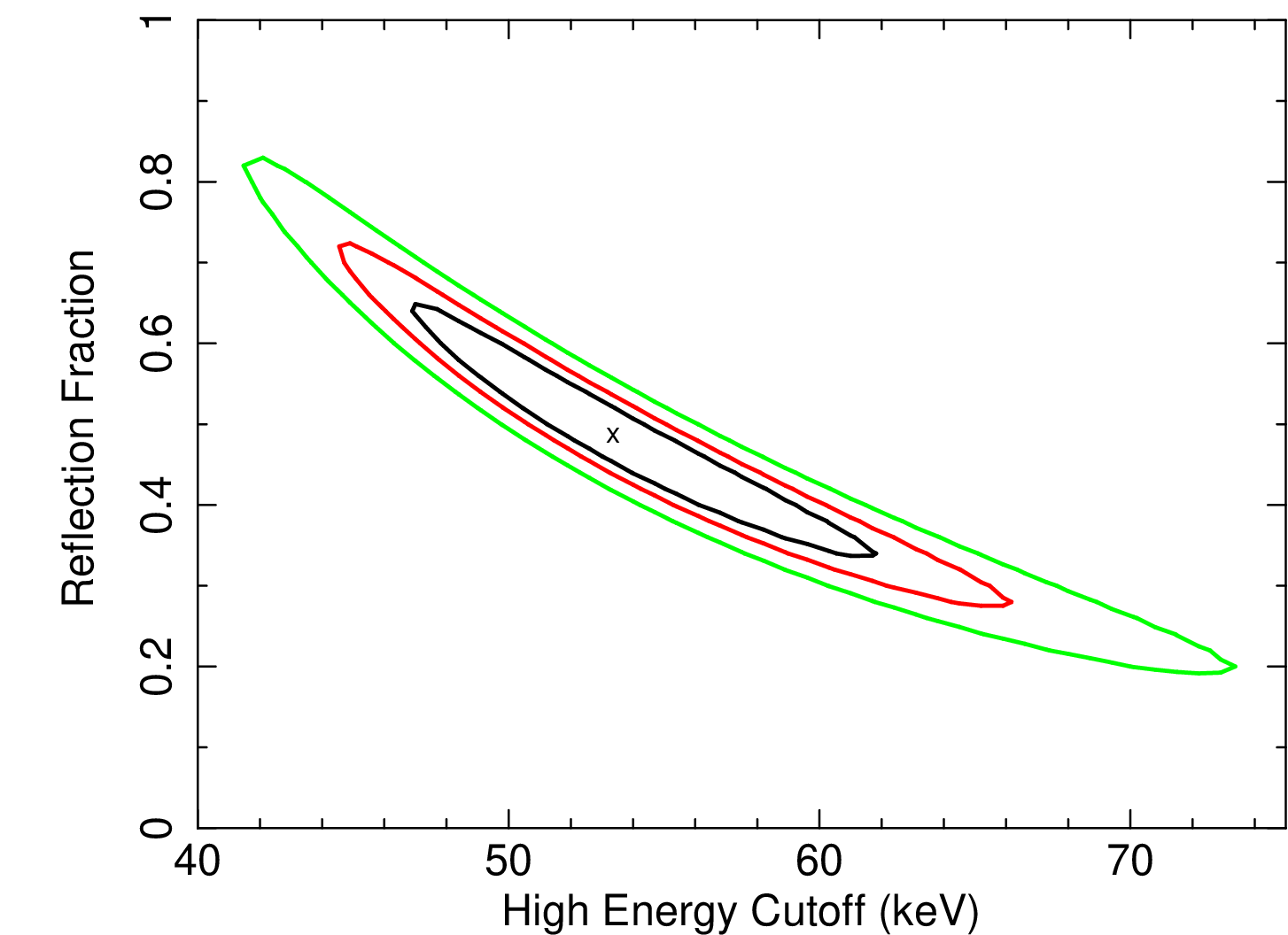}
 \caption{$E_{c}$-$\Gamma$ contour plot (left panel) and $E_{c}$-$R$ contour plot (right panel) for the \textit{NuSTAR} observation. The solid black, red and green curves refer to the $68$, $90$ and $99\%$ confidence levels, respectively. The X represents the best fit value of the parameters.}
\label{fig_cont}
\end{figure*}
The fit shows a weaker iron line with respect to what we expected from the Compton hump. Replacing the \textsc{pexrav} model with a self-consistent model that includes the Fe K$\alpha$ line, such as the \textsc{pexmon} model\footnote{\textsc{xspec} model: constant * TBabs * (mekal + mtable$\lbrace$cloudy.fits$\rbrace$ * zwabs * pexmon)} (\cite{nandra}), with the relative iron abundance left free to vary, a value of $0.6 \pm 0.3$ is
found for this parameter ($\chi^2$=$585$ for $542$ d.o.f.)\\
In order to test for the presence of a relativistic component, we fitted the data with the \textsc{relxill} model\footnote{\textsc{xspec} model: constant * TBabs * (mekal + mtable$\lbrace$cloudy.fits$\rbrace$ * zwabs * relxill)} (\cite{garcia2014}). Since the black hole spin parameter was not constrained, we assumed $a=0.998$. We fixed the reflection fraction parameter to the best fit values found with the previous best fit model (see Table \ref{fit_pexrav}). Including the relativistic effects provides no improvement in the fit, implying that no relativistic component is required by the data.\\
\subsection{Comptonization features}
Finally, assuming that the primary emission is due to Comptonization of thermal disc photons in a hot corona, we estimated the coronal parameters using an analytical Comptonization model. The temperature is expected to be related to the cutoff energy by $E_c=2-3 \times kT_{\rm e}$ (\cite{Petrucci2000}, \cite{Petrucci2001}), so, for such a low value of the cutoff energy ($53^{+11}_{-8}$ keV), we expect a low value for the coronal temperature, and a high value for the optical depth to account for the flat spectrum. We fitted the \textit{NuSTAR} spectra with the \textsc{comptt} model (\cite{titcomptt}), adding the reflection component computed by \textsc{pexrav} with two Gaussian lines\footnote{\textsc{xspec} model: constant * TBabs * zwabs * (compTT + zgauss + zgauss + pexrav)}(the iron K$\alpha$ emission line and the absorption line due to Fe\textsc{XXV} K$\alpha$ ions). Because of the low \textit{NuSTAR} spectral resolution we fixed the centroid energies of the lines to the values found in the best fit of the \textit{XMM-Newton} data. In this model the seed photon spectrum is a Wien law; we fixed the temperature to the maximum temperature of the accretion disk, which in this case, for \cite{shakura} disk is $4$ eV, given the black hole mass of $\sim3 \times 10^8$ solar masses (see the Appendix). In the case of a slab geometry of the corona, we found a coronal temperature kT$_{\rm e}=12.1^{+1.8}_{-1.2}$ keV and an optical depth $\tau=2.8^{+0.2}_{-0.3}$. The fit is good, with a $\chi^2$ of $411$ for $383$ d.o.f.. For the case of a spherical geometry, we found a statistically equivalent fit. The value of the coronal temperature is about the same, while the optical depth is higher by almost a factor of two: $\tau=6.3^{+0.4}_{-0.5}$. The difference is primarily due to the different meaning of this parameter in the two geometries: the optical depth for a slab geometry is the average of optical depth values along the different directions, so it is lower than the effective value, while that for a sphere is the radial one (see \cite{titcomptt} for a more detailed description). 
\begin{figure*}
\includegraphics[width=0.7\columnwidth, angle=-90]{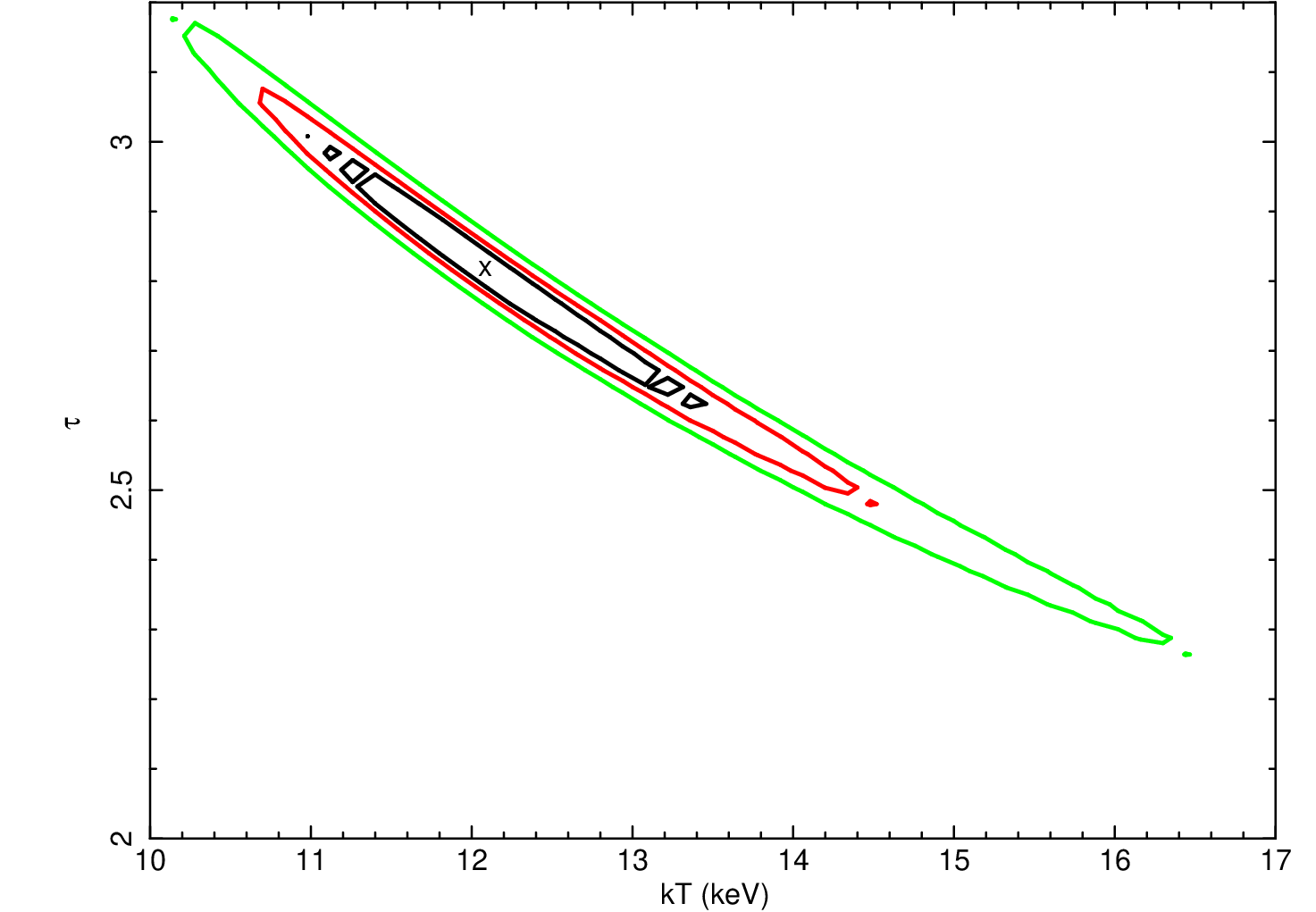}
\includegraphics[width=0.7\columnwidth, angle=-90]{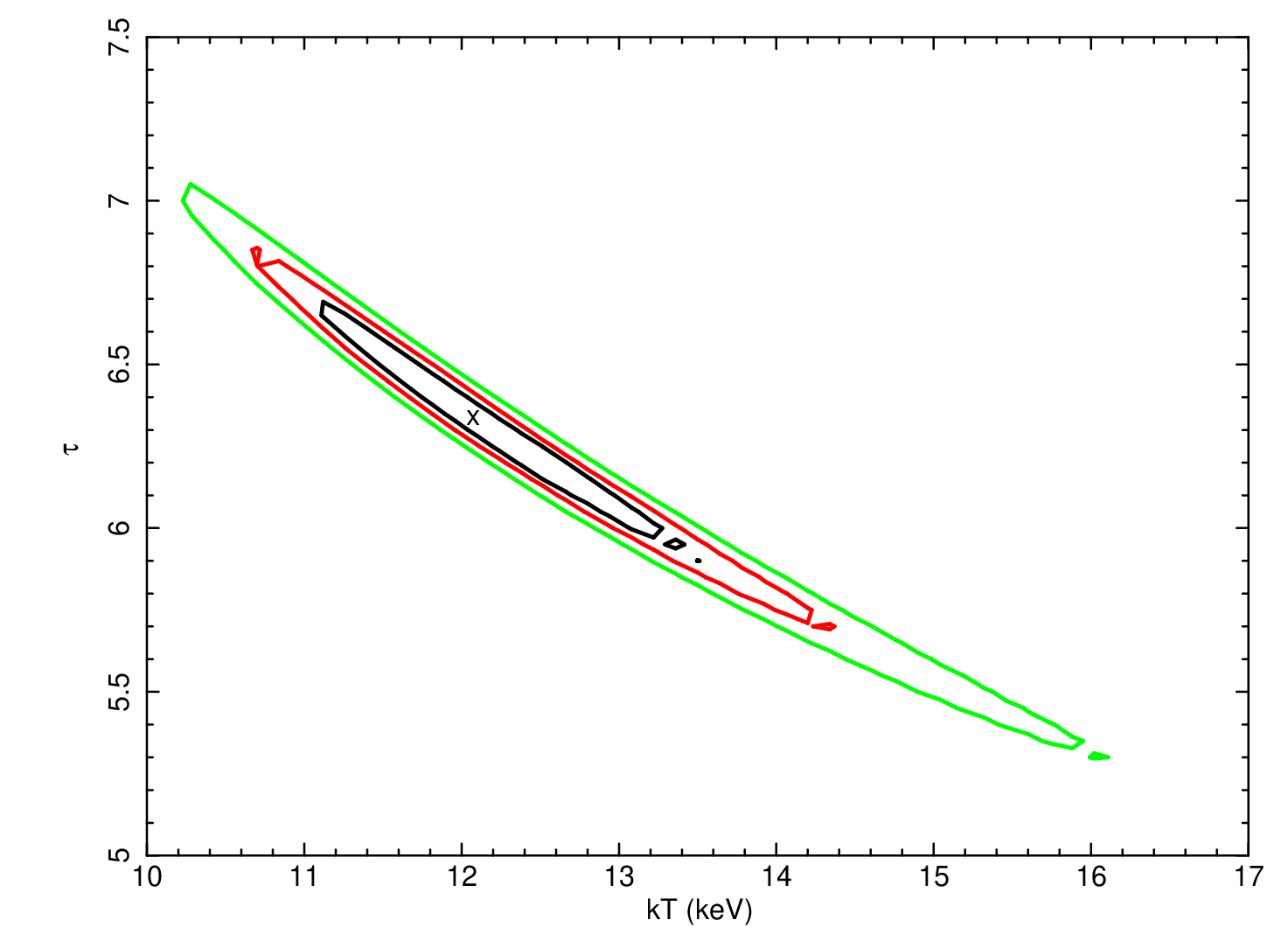}
 \caption{Coronal temperature vs optical depth contour plot in the case of slab geometry (left panel) and spherical geometry (right panel) for the \textit{NuSTAR} observation when the \textsc{comptt} model is used to fit the data. The solid black, red and green curves refer to the $68$, $90$ and $99\%$ confidence levels respectively. The X represents the best fit value of the parameters.}
\label{compttcont}
\end{figure*}
The contour plots of the coronal temperature versus the optical depth obtained with the two different geometries are shown in Figure \ref{compttcont}.\\
We did not try to fit the Comptonization with the \textsc{compps} model because the optical depth values obtained with the \textsc{comptt} model are too high and they do not fall within the region of parameter space where the numerical \textsc{compps} method produces reasonable results (see \cite{PoutanenSvensson} for more details).\\

\section{Discussion and Conclusions}
\label{concl}

We have presented an analysis of non-simultaneous \textit{XMM-Newton} and \textit{NuSTAR} observations of the Seyfert 1 galaxy GRS 1734-292. The spectral slope of the primary power law is different between the two observations, being very flat in the \textit{XMM-Newton} observation ($\Gamma\sim$1.47, consistent with the values found by \cite{guainazzi2011}) while it is more typical of a Seyfert galaxy in the \textit{NuSTAR} observation ($\Gamma\sim$1.65), when the source was a factor of $\sim 1.3$ brighter. The variation of the spectral photon index could be
associated with a variation of the coronal parameters and in particular of the optical depth.\\
The $2-10$ keV absorption-corrected luminosity from the \textit{XMM-Newton} observation is $L_{2-10}= 5.23 \pm 0.03 \times 10^{43}$ erg s$^{-1}$. 
Using the $2-10$ keV bolometric correction of \cite{Marconi2004}, we estimate the bolometric luminosity to be $L_{\rm bol}=1.45 \times 10^{45}$erg s$^{-1}$. From the bolometric luminosity, with the black hole mass as in Appendix \ref{optical}, we estimate the $L_{\rm bol}/L_{\rm Edd}$ ratio to be $0.033$. \\
The presence of an iron K$\alpha$ emission line at $6.4$ keV, albeit weak, is confirmed. We found also one absorption line, with a  centroid energy at around $6.69$ keV, which is consistent with the energy expected for the K-shell transition of Fe\textsc{XXV} ions. The cutoff energy is $53^{+11}_{-8}$ keV, fully consistent with that found by \cite{Malizia14}. This is the lowest value found so far by \textit{NuSTAR} in a Seyfert galaxy together with Mrk 335 (\cite{keek16}); comparable or even lower values are found in stellar-mass accreting black holes (\cite{miller13}, \cite{miller15}). We estimated the coronal parameters by fitting the \textit{NuSTAR} data with the \textsc{comptt} Comptonization model, finding a coronal temperature of $kT_e=12.1^{+1.8}_{-1.28}$ keV and an optical depth $\tau=2.8^{+0.2}_{-0.3}$ assuming a slab geometry or a similar temperature and  $\tau=6.38^{+0.4}_{-0.5}$ assuming a spherical geometry. Of course, we are implicitely assuming a simple picture in which the corona is a single temperature zone, which may not be the case if the heating is localized, as e.g. in the case of magnetic reconnection. \\
We used these values to put GRS 1734-292 in the compactness-temperature ($\Theta_e$ - $\ell$) diagram (\cite{Fabian}, and references therein). Here $\Theta_e=kT_e/m_ec^2$ is the electron temperature normalized to the electron rest energy and $\ell$ is the dimensionless compactness parameter \citep{Fabian}:
\begin{equation}
\ell=\frac{L}{R}\frac{\sigma_T}{m_ec^3}
\end{equation}
 where $L$ is the luminosity and $R$ is the radius of the corona (assumed spherical). We obtain $\Theta_e = 0.023^{+0.004}_{-0.002}$. To compute the compactness parameter, following \cite{Fabian05} we adopted the luminosity of the power-law component extrapolated to the $0.1-200$ keV band; since no measurement exists for the radius, we assume a value of $10$ gravitational radii $R_g$. We found $\ell=13.3 \pm 0.3 (R_{10})^{-1}$ were $R_{10}$ is the ratio between the radius and $10 R_g$.\\ As obvious, given the low coronal temperature, GRS 1734-292 is located far away from the region of pair production in the $\Theta_e$ - $\ell$ plane, and is also located well below the $e^--e^-$ coupling line (i.e. the line below which the electron-electron coupling time scale is shorter than the Compton cooling time scale). This should ensure that the electron population is thermalized. It is instead located close to the $e^--p$ coupling line, below which the electron-proton coupling time scale is shorter than the Compton cooling time scale. It is interesting to note that no sources among those analized by \cite{Fabian05} lie definitely below the $e^--p$ line, while a number of them lie around or just above (see Fig. 4 in their paper). This line therefore seems to set a physical boundary, which may be understood, at least qualitatively, noting that if the electron population cools by Compton scattering and its temperature decreases until electron-proton coupling becomes important, and the transfer of energy from protons to electrons effective. This is not a completely self-consistent picture, as the electron-proton coupling line was calculated assuming that the electron and proton temperatures (normalized to their mass), $\Theta_e$ and $\Theta_p$, are the same (\cite{Fabian94}), which is unlikely when Compton cooling dominates. Moreover, the dependence of the coupling time on $\Theta_e$ is small as soon as the two temperatures are decoupled and the proton temperature is the largest. Time-dependent, detailed calculations with realistic heating and energy redistribution mechanisms are required to assess how effective this feedback may be.\\
Only a few AGN in the \cite{Fabian05} compilation have temperatures as low as that of GRS 1734-292, and none among those observed by \textit{NuSTAR}. We note that the accretion rate of GRS 1734-292 is only a few percent of the Eddington limit, so the effectiveness of the cooling mechanism cannot be related to a particuarly strong radiation field. It may, however, be at least partly related to the high value of the optical depth $\tau$. A seed photon coming from the disc, in fact, will undergo more than one scattering before leaving the corona, thereby reducing the electron temperature. Indeed, models predict an anticorrelation between coronal temperature and optical depth (see e.g. \cite{Petrucci2001} for a calculation based on the two-phase model of \cite{HeM93}: note that values not too different from ours are predicted). The reason for the unusually large value of the optical depth is unclear (but see \cite{keek16} for evidence of an increase of the optical depth with decreasing Eddington ratio in Mrk 335), and difficult to assess given our poor knowledge of the processes which originate the corona and of the mechanisms which transfer the energy there. But with the increasing amount of high quality spectra from \textit{NuSTAR}, progressively populating this parameter space, it is at least possible to start seriously pondering these questions.

\section*{Acknowledgements}
We thank the anonymous referee for comments which helped improving the clarity of the paper. This work made use of data from the \textit{NuSTAR} mission, a project led by the California Institute of Technology, managed by the Jet Propulsion Laboratory, and funded by the National Aeronautics and Space Administration. We thank the \textit{NuSTAR} Operations, Software and Calibration teams for support with the execution and analysis of these observations. This research has made use of the\textit{ NuSTAR} Data Analysis Software (NuSTARDAS) jointly developed by the ASI Science Data Center (ASDC, Italy) and the California Institute of Technology (USA). The work is also based on observations obtained with XMM--$Newton$, an ESA science mission with instruments and contributions directly funded by ESA Member States and the USA (NASA). AT, AM, GM and FU acknowledge financial support from Italian Space Agency under grant  ASI/INAF I/037/12/0-011/13, SB under grant ASI-INAF I/037/12/P1. AT, AM, SB and GM acknowledge financial support from the European Union Seventh Framework Programme (FP7/2007-2013) under grant agreement no. 312789. 





\appendix
\section{black hole mass estimate.}
\label{optical}
The spectroscopic observation of the optical counterpart of GRS 1734-292 was carried out with the EFOSC2 instrument mounted on the 3.6m ESO-NTT telescope at La Silla, on 2010-07-08 (program ID:085.D-0441(C), PI: Jonker), using GRISM 13 and a 1'' slit. The pointing is 500 s long and we used \textsc{iraf} (version 2.16) and \textsc{midas} (release 15SEPpl1.0) for data reduction and calibration, using standard procedures.

The aim of our analysis was to measure the width of the broad H$\alpha$ $\lambda6563$ component, to infer the black hole mass via a virial-based, single-epoch relation (\cite{LaFranca15}, \cite{Ricci16}). In Fig. A1 and A2 the 4700-7500 $\angstrom$ and 6500-7500 $\angstrom$ spectra of GRS 1734-292, respectively, are shown: several emission lines of H, O, N and S elements can be clearly seen. Throughout our analysis we assumed that F([N {\sc ii}] $\lambda6583$)/F([N {\sc ii}] $\lambda6548$)=3, as required by the ratio of the respective Einstein coefficients. Spectra are fitted with {\sc xspec}, via $\chi^2$ minimization, by modelling the continuum as a power law convolved with a {\sc spline} function, and each line component as a Gaussian. The width of the narrow lines was fixed to the instrumental one, inferred from fitting the He-Ar calibration lines. We assumed a redshift z=0.0214 (\cite{Marti1998}) and all reported wavelenghts in Table \ref{t_opt} are rest-frame The inferred fluxes for the H$\alpha$ (broad component) and H$\beta$ emission lines lead to an observed H$\alpha$/H$\beta>12.6$. Assuming an average Balmer-line intensity relative to H$\beta$ of 2.86 (case B recombination) we calculate a Galactic extinction in the V band of A$_V>4.6$ mag. Adopting the standard  Galactic gas-to-dust ratio, the optical reddening may be rewritten using the relation A$_V = 5.27 N^{22}_ H$mag, where the absorbing column density is expressed in units of  $10^{22}$ cm$^{−2}$ (see e.g. \cite{maiolino}, and references therein). The lower limit obtained with the optical data analysis is in agreement with the absorbing column density measured from the X-ray spectrum.
\begin{figure}
\centering
\includegraphics[width=0.7\columnwidth, angle=-90]{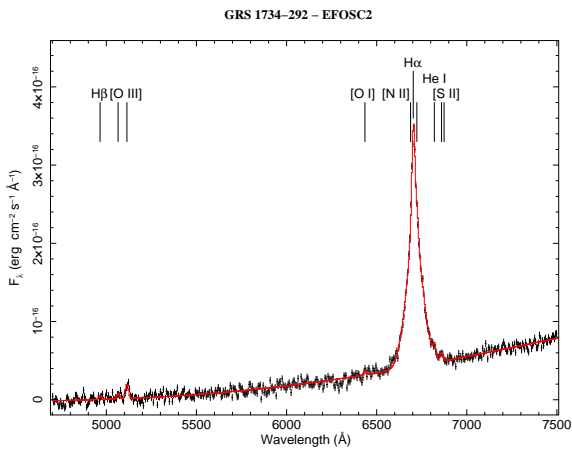}
\caption{ESO-NTT optical spectrum of the source, in the 4700-7500 $\angstrom$ range. Emission lines from several elements such as H, O, N and S are clearly detected.}
\end{figure}

\begin{table}
 \centering
 \caption{Optical emission lines in the ESO-NTT spectrum of GRS 1734-292.}
\begin{tabular}{lccc}
\hline
 Line & $\lambda$ & Flux & FWHM  \\
(1) & (2) & (3) &(4) \\
\hline
& & & \\
H$\beta$ & 4861.33 & $<1.5$ & -\\
$[\rm{O}\ \textsc{iii}]$ &4958.92 &$0.13\pm0.08$& - \\
$[\rm{O}\ \textsc{iii}]$ & 5006.85 & $0.38\pm0.08$& -\\
$[\rm{O}\ \textsc{i}]$ & 6300.32& $0.10\pm0.08$& -\\
$[\rm{N}\ \textsc{ii}]$ & 6548.06& $0.30\pm0.03$& -\\
 H$\alpha$ Nr.&6562.79 & $3.02^{+0.06}_{-0.13}$& -\\
 H$\alpha$ Br.&6562.79 & $19.03^{+0.26}_{-0.20}$&$4940\pm50$ \\
$[\rm{N}\ \textsc{ii}]$ &6583.39 & $0.90\pm0.09$& -\\
$\rm{He}\ \textsc{i}$ &  6678.11 & $0.3 \pm 0.1$ & - \\
$[\rm{S}\ \textsc{ii}]$ & 6716.42& $0.20\pm0.07$& -\\
$[\rm{S}\ \textsc{ii}]$ & 6730.78& $0.15 \pm 0.05$& -\\
\hline
\end{tabular} \\
\label{t_opt}
Notes. Col. (1) Identification. (2) Laboratory wavelength ($\angstrom$) (air: Bowen 1960). (3)FWHM in km s$^{-1}$ units. Dashes indicate a fixed FWHM=18 $\angstrom$. Col (4) Fluxes in 10$^{-15}$ erg cm$^{-2}$ s$^{-1}$ units.
\end{table}

We measured a FWHM=$4940\pm50$ km s$^{-1}$ for the broad component of the H$\alpha$ line. This value and the 2-10 keV luminosity measured with XMM-{\it Newton}, which is the closest observation in time (L$_{\rm X}=5.23\pm0.03\times 10^{43}$ erg s$^{-1}$), allow us to use the updated calibrations of the virial black hole mass estimators (\cite{Ricci16}). The inferred mass is $\log(M_{\rm bh}/M_{\odot})=8.5$, with an intrinsic spread of the relation of $\sim0.5$ dex.

\begin{figure}
\centering
\includegraphics[width=0.7\columnwidth, angle=-90]{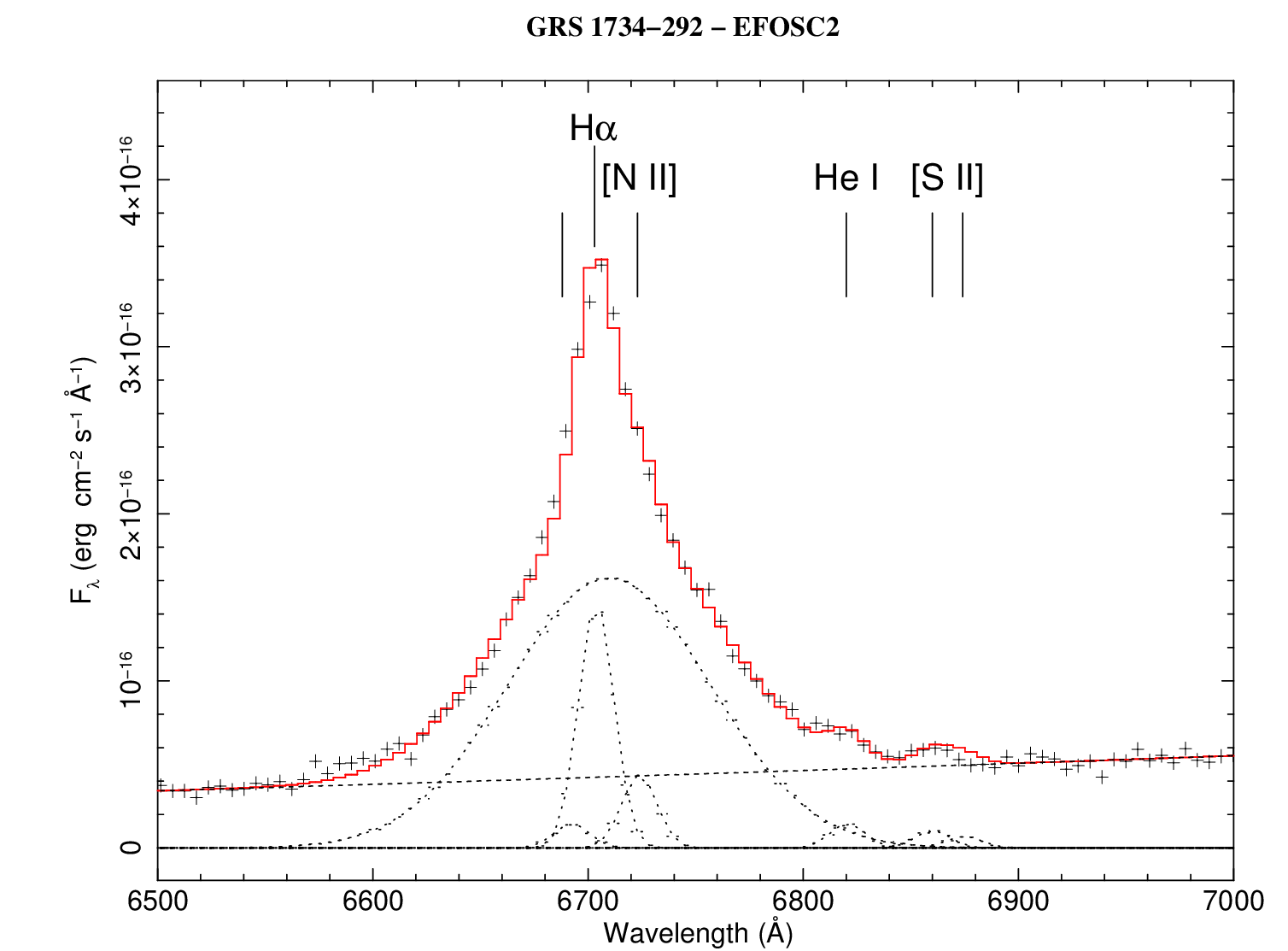}
\caption{ESO-NTT optical spectrum of the source, in the 6500-7500 $\angstrom$ range.}
\end{figure}


\bsp	
\label{lastpage}
\end{document}